\documentclass[doublecol]{epl2} 

\usepackage{amsmath,amssymb}
\usepackage{bm}
\usepackage{physics}

\usepackage{hyperref}

\newcommand\eq{\text{eq}}
\newcommand\st{\text{s}}
\newcommand\hcs{\text{HCS}}
\newcommand{\sgn}{\text{sgn}}

\title{Non-equilibrium memory effects: granular fluids and beyond}

\author{A. Patrón\inst{1}, B. Sánchez-Rey\inst{2}, C. A. Plata\inst{1}, and A. Prados\inst{1}}

\institute{                    
  \inst{1} Física Teórica, Universidad de Sevilla, Apartado de
  Correos 1065, E-41080 Sevilla, Spain\\
  \inst{2} Departamento de Física Aplicada I, E.P.S., Universidad de
  Sevilla, Virgen de África 7, E-41011 Sevilla, Spain
  }

\abstract{ In this perspective paper, we look into memory effects in
  out-of-equilibrium systems. To be concrete, we exemplify memory
  effects with the paradigmatic case of granular fluids, although
  extensions to other contexts such as molecular fluids with non-linear drag are also considered. The focus is put on two archetypal memory effects: the Kovacs and Mpemba effects. In brief, the first is related to imperfectly reaching a steady state---either equilibrium or
  non-equilibrium, whereas the second is related to reaching a steady
  state faster despite starting further. Connections to optimal control theory thus naturally emerge and are briefly discussed.}

\begin{document}

\maketitle

\section{Introduction}\label{sec:intro}

Under quite general conditions, many physical systems tend in the long
time limit to a state in which all trace of initial conditions is
lost. This state is often stationary, either an equilibrium state or a
non-equilibrium steady state (NESS), but it also may be a
time-dependent ``hydrodynamic'' state---in which a reduced description in terms of a few ``thermodynamic'' or
``macrosocopic'' variables accounts for the complete characterisation of the time evolution of the system.

Memory effects are intimately related to
aging~\cite{cugliandolo_evidence_1994,keim_memory_2019,jaeger_temperature_2022}. A
system displays aging when its relaxation or time correlations are not
invariant under time translation after being aged for a long waiting time; instead, they explicitly depend on such a  time. A memory effect emerges in a physical system when its time
evolution depends on the previous history, i.e.~on its initial
preparation that, in turn, depends on how it has been previously
``aged''.

A classic example of memory effect is the so-called Kovacs hump,
first reported by Kovacs for the  volume relaxation of polymeric
glasses~\cite{kovacs_transition_1963,kovacs_isobaric_1979}. Analogous
behaviours have been repeatedly observed in different
contexts~\cite{chow_molecular_1983,berthier_surfing_2002,buhot_kovacs_2003,bertin_kovacs_2003,arenzon_kovacs_2004,cugliandolo_memory_2004,mossa_crossover_2004,aquino_kovacs_2006,prados_kovacs_2010,bouchbinder_nonequilibrium_2010,diezemann_memory_2011,chang_kovacs_2013,ruiz-garcia_kovacs_2014,prados_kovacs-like_2014,trizac_memory_2014,brey_memory_2014,
  plata_kovacs-like_2017,kursten_giant_2017,  mompo_memory_2021,peyrard_memory_2020,mandal_memory_2021,militaru_kovacs_2021,lulli_kovacs_2021,sanchez-rey_linear_2021,patron_strong_2021,godreche_glauber-ising_2022}.
Let us consider a 
quantity $P$ of a physical system in contact with a thermal bath.  Its
equilibrium value is denoted by $P_{\eq}(T)$, which is assumed to be a monotonic function. The Kovacs hump is the non-monotonic response
of the system to the two-jump protocol described below.

\begin{figure}
\centering
\includegraphics[width=2.75in]{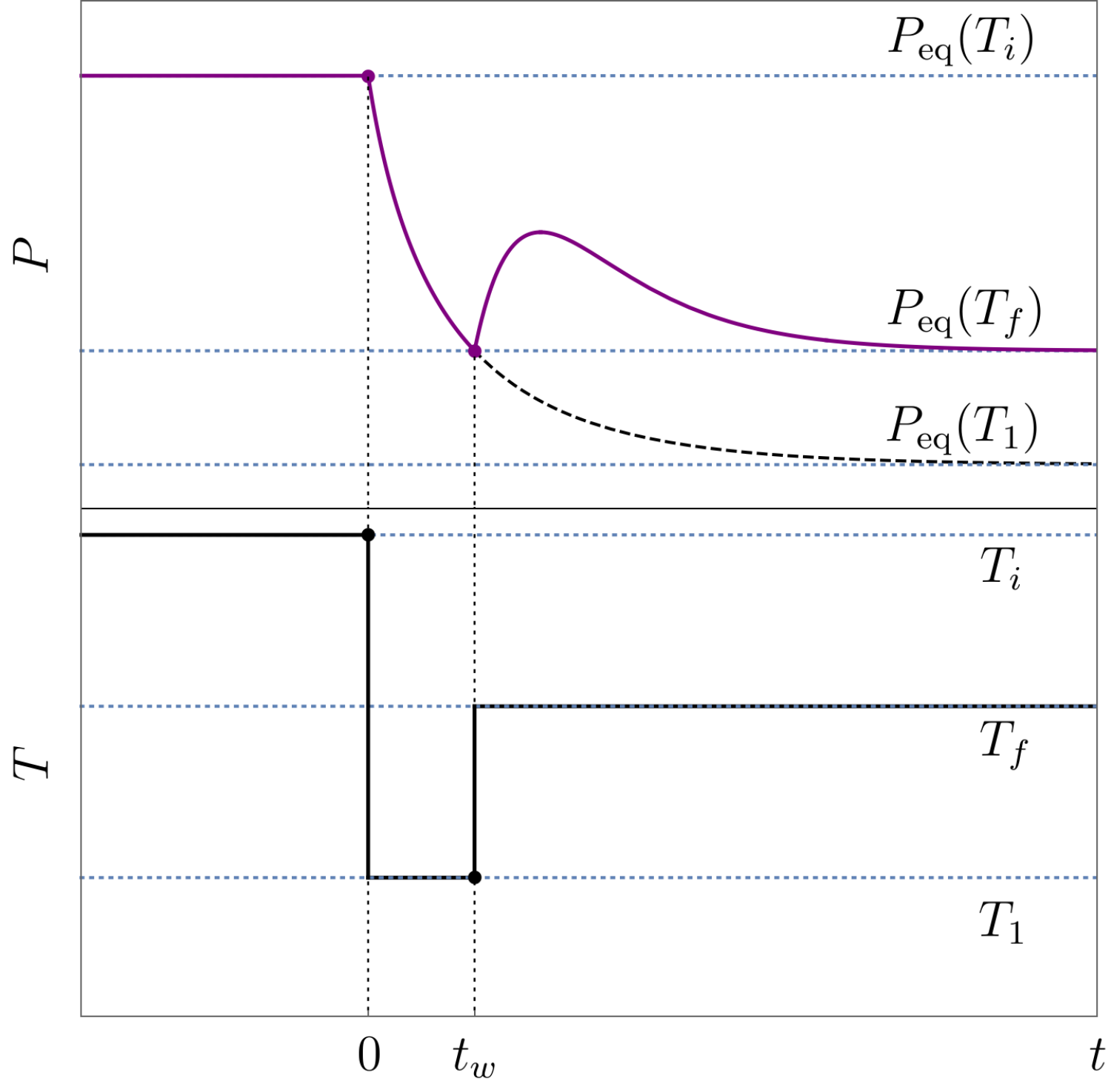}
\caption{Qualitative picture of the Kovacs hump. The time evolution of a physical quantity $P$ is depicted on the top panel, when the system is submitted to the two-jump protocol in the temperature shown on the bottom panel. The relaxation from $T_i$ to $T_1$ (dashed line) is interrupted at $t=t_w$, when the quantity $P$ has its equilibrium value at the final temperature $T_{f}$, $P(t_w)=P_{\eq}(T_{f})$. Nevertheless, $P(t)$ deviates from $P_{\eq}(T_{f})$ and passes through a maximum before returning thereto---thus showing the need of additional physical quantities to completely characterise the state of the system.}
\label{fig:kovacs}
\end{figure}
Figure~\ref{fig:kovacs} shows a sketch of the Kovacs protocol and the associated Kovacs response. The system is initially
equilibrated at temperature $T_{i}$, and therefrom aged at a
lower temperature $T_{1}<T_{i}$ in the time interval $0<t<t_{w}$. At
$t=t_{w}$, the instantaneous value of $P$ is $P(t_{w})$, and thereat the temperature of the bath is abruptly changed to $T_{f}$, with $P(t_{w})=P_{\eq}(T_{f})$---thus, $T_{i}>T_{f}>T_{1}$. The system
displays the Kovacs effect when, for $t>t_{w}$, $P$ departs from its equilibrium
value, which $P$ has as a consequence of the choice of
$T_{f}$, and presents a non-monotonic behaviour. The existence of this Kovacs hump entails that the pair
$(T,P)$ does not suffice to completely characterise the state of the
system: additional state variables are necessary.

Another example of memory effect is the Mpemba
effect~\cite{mpemba_cool_1969}. Originally, the
Mpemba effect refers to ``hot'' water freezing faster than ``cold''
water~\cite{mpemba_cool_1969,jin_mechanisms_2015}{, in contradiction with the usual Newton's law of cooling~\cite{maruyama_newtons_2021}}. In this context, the very existence of
the Mpemba effect is still
controversial~\cite{burridge_questioning_2016,burridge_observing_2020}. Recently, the Mpemba effect has attracted the attention of the
non-equilibrium physics community, understanding it in a generalised
way as follows. The relaxation of two samples of the same system
to a common final steady state is considered. Under certain
conditions, the sample initially further from the steady
state relaxes thereto faster than that initially closer. 

\begin{figure}
\centering
\includegraphics[width=2.75in]{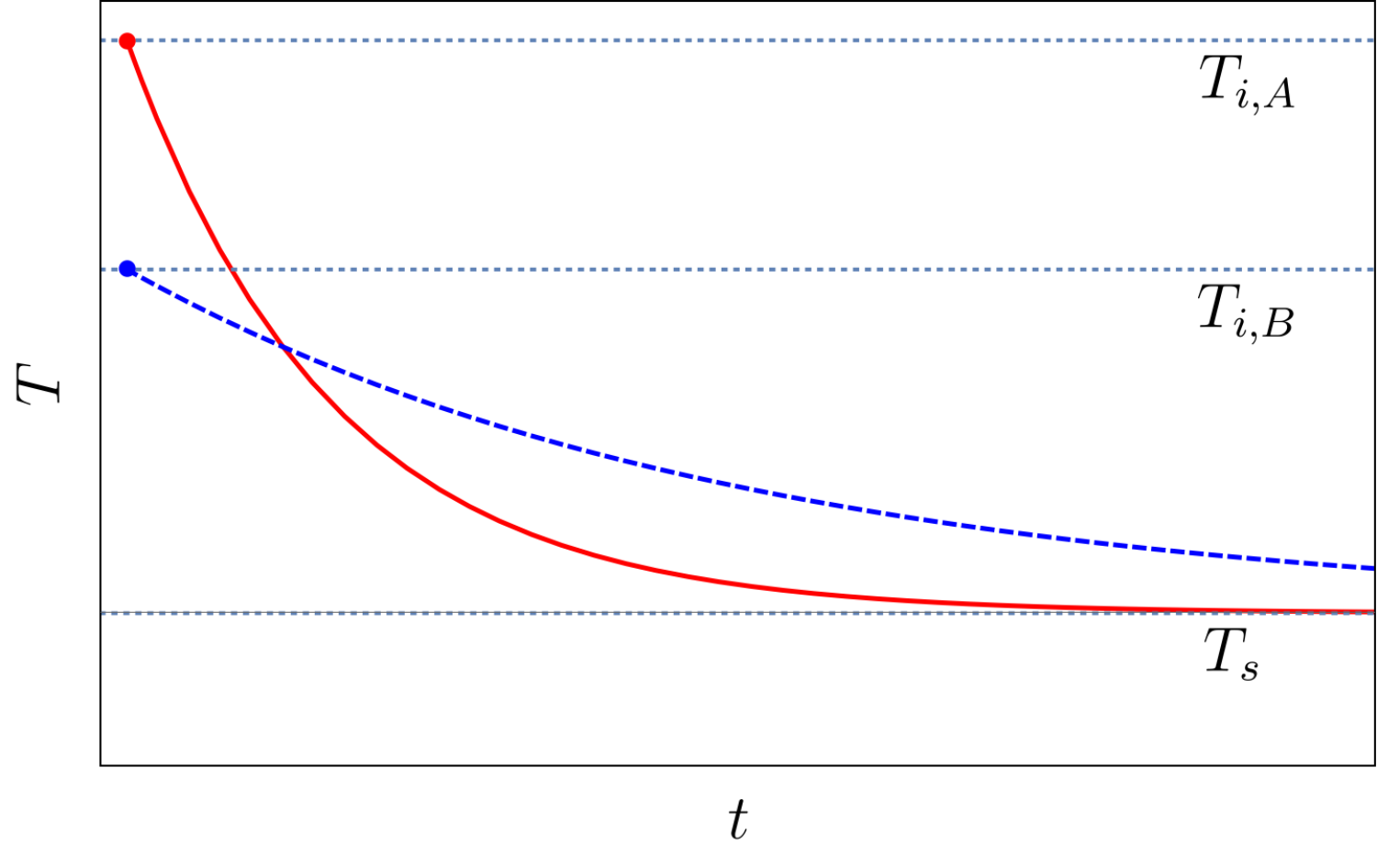}
\caption{Qualitative picture of the Mpemba memory effect. The ``hot" sample $A$, with initial kinetic temperature $T_{i,A}$, is further from equilibrium at the common bath temperature $T_s$ than the ``cold" sample $B$, with initial temperature $T_{i,B}<T_{i,A}$.  {In the kinetic approach,} the thermal Mpemba effect emerges when the time evolution of the initially hotter sample (red solid line) overtakes that of the initially colder one (blue dashed). In the stochastic processes approach, the evolution of the distance in probability space---e.g. the Kullback-Leibler divergence---is monitored instead of the kinetic temperature, and the entropic Mpemba effect arises when a similar crossing of the relaxation towards equilibrium of samples $A$ and $B$ is observed. }
\label{fig:mpemba}
\end{figure}

The Mpemba effect is qualitatively depicted in fig.~\ref{fig:mpemba}. Both the Mpemba---the hotter cools
sooner---and the inverse Mpemba---the colder heats sooner---effects have
been observed in many different systems~\cite{lu_nonequilibrium_2017,lasanta_when_2017,klich_mpemba_2019,torrente_large_2019,baity-jesi_mpemba_2019,yang_non-markovian_2020,biswas_mpemba_2020,mompo_memory_2021,santos_mpemba_2020,carollo_exponentially_2021,chetrite_metastable_2021,uskokovic_and_2021,biswas_mpemba_2021,busiello_inducing_2021,takada_mpemba_2021,gomez_gonzalez_mpemba-like_2021,
  patron_strong_2021,megias_mpemba-like_2022,zhang_theoretical_2022,degunther_anomalous_2022,biswas_mpemba_2022,lin_power_2022,megias_thermal_2022,holtzman_landau_2022,kumar_anomalous_2022,yang_mpemba_2022,chorazewski_curious_2023,biswas_mpemba_2023,sun_physics_2023,chatterjee_quantum_2023,teza_relaxation_2023}. In
the theoretical studies, two main approaches have
been used: stochastic (entropic)~\cite{lu_nonequilibrium_2017} and kinetic (thermal)~\cite{lasanta_when_2017}, which we describe later in detail. In the former, distance to equilibrium is defined in probability space, e.g.\ with the Kullback-Leibler divergence. In the latter, distance to equilibrium is monitored through the kinetic temperature, which is proportional to the average kinetic energy.\footnote{For equilibrium systems, due to the equipartition theorem, the
kinetic temperature equals the thermodynamic temperature. This is no longer the case for out-of-equilibrium states.}

In the Mpemba effect, the system that is further from the steady state  somehow takes a shortcut and thus relaxes thereto faster than the closer one. Then,
there appears a natural connection with the general field of shortcuts
or, employing the terminology introduced in
ref.~\cite{guery-odelin_driving_2023}, swift state to state
transformations. In particular, a related problem is the optimisation
of the relaxation route to equilibrium---or to a NESS. For given
initial and final states, the minimisation of the connection time
between them by engineering the time dependence of some physical
quantities, like the temperature or the potential, is a well-defined
mathematical problem in optimal control
theory~\cite{liberzon_calculus_2012}. This
is the classic brachistochrone problem, which very recently has been
addressed for both quantum and non-equilibrium
systems~\cite{deffner_quantum_2017,plata_finite-time_2020,lam_demonstration_2021,prados_optimizing_2021,ruiz-pino_optimal_2022,patron_thermal_2022,guery-odelin_driving_2023,aghion_thermodynamic_2023,pires_optimal_2023}.


\section{Kovacs effect}\label{sec:Kovacs}

For systems with a master equation dynamics, there are
general results for the shape of the Kovacs hump in linear
response. These results hold under quite general conditions, basically
(i) a canonical form of the equilibrium probability distribution function (pdf), proportional to
$\exp(-\beta H)$, with $\beta=(k_{B}T)^{-1}$ and $H$ being the
system's Hamiltonian, and (ii) detailed
balance in the dynamics~\cite{prados_kovacs_2010}. With these assumptions, the form of
the Kovacs hump for the energy $E(t)=\expval{H}(t)$ is directly related to
the form of its ``direct'' relaxation function $\phi_{E}(t)$ from $T_{i}$
to $T_{f}$, with only one jump. 

From the explicit expression of the Kovacs hump in linear response, eq.~(43)
of ref.~\cite{prados_kovacs_2010}, one deduces that: (i) the Kovacs
hump is always positive, i.e. $E(t)\ge E_{\eq}(T_{f})$, (ii) there is
only one maximum of $E(t)$. Interestingly, the explicit
  expression of the Kovacs hump derived in
  ref.~\cite{prados_kovacs_2010} resembles the phenomenological
  expression written by Kovacs~\cite{kovacs_isobaric_1979}. Although
the majority of studies are done in the non-linear regime, i.e. with
large values of the temperature jumps, the behaviour described by the linear response theory, i.e. (i)
and (ii) above, a positive hump with only one maximum, is the one
found in glassy and other complex
systems~\cite{kovacs_transition_1963,kovacs_isobaric_1979,chow_molecular_1983,berthier_surfing_2002,buhot_kovacs_2003,bertin_kovacs_2003,cugliandolo_memory_2004,arenzon_kovacs_2004,mossa_crossover_2004,aquino_kovacs_2006,prados_kovacs_2010,bouchbinder_nonequilibrium_2010,diezemann_memory_2011,chang_kovacs_2013,ruiz-garcia_kovacs_2014,peyrard_memory_2020,mandal_memory_2021,
lulli_kovacs_2021,patron_strong_2021,godreche_glauber-ising_2022};
thus the term ``normal Kovacs hump'' has been coined to describe
it. The normal hump stems from the structure of the direct relaxation
function in linear response, which is a sum of exponentially
decreasing modes with positive coefficients.


In glassy systems, the emergence of the Kovacs effect is often
explained as a consequence of the complex energy landscape typical
thereof. Still, the Kovacs effect has also been observed in systems
with a much simpler energy landscape. A paradigmatic case is that of
granular gases, which are intrinsically non-equilibrium systems:
energy is purely kinetic but it is continuously dissipated in
collisions. 
Therefore, an external mechanism is needed to drive the
system to a stationary state, which is always a NESS with a
non-Maxwellian velocity distribution function (vdf)~\cite{poschel_granular_2001}. The simplest one is that of the uniformly heated granular gas, in
which independent white noise forces with variance $\chi$ act on all
 particles. The kinetic temperature---here also called granular temperature---at the NESS is then a certain
function of $\chi$~\cite{van_noije_velocity_1998,montanero_computer_2000}. 

Despite the simple energy landscape, the Kovacs effect neatly appears
when a system of smooth inelastic
  hard
particles\footnote{In smooth collisions, the tangential component of the relative velocity is conserved, whereas the normal component is reversed and shrunk with the restitution coefficient $\alpha$; the energy loss is thus proportional to $1-\alpha^2$ and $\alpha=1$ corresponds to the elastic case.} is submitted to the two-jump Kovacs protocol, with the
intensity of the driving playing the role of the bath temperature,
$\chi_{i}\to \chi_{1} \to
\chi_{f}$~\cite{prados_kovacs-like_2014,trizac_memory_2014}. This
entails that the instantaneous value of the kinetic temperature
$T(t)$ does not suffice to completely describe granular fluids. It is
the non-Gaussianities that are responsible for the emergence of the
Kovacs effect, and it is thus essential to incorporate them to the
physical picture. It suffices to do so in the simplest way by
including only the excess kurtosis $a_{2}(t)$---the
so-called first Sonine approximation.

More interestingly, the sign of the Kovacs hump depends on the
inelasticity. Specifically, it depends on the sign of the excess
kurtosis at the steady state, $a_{2}^{\st}$, which is negative
(positive) for small (large) inelasticity. The key point is the
cooling rate being an increasing function of $a_{2}$. By aging the
system with a very low value of the driving $\chi_{1}$, the system
falls onto the homogeneous cooling state (HCS)~\cite{brey_homogeneous_1996},
in which the granular fluid freely cools following Haff's law~\cite{haff_grain_1983}, $T(t)\propto t^{-2}$, and the excess kurtosis becomes constant and equals $a_{2}^{\hcs}$. One always has $\sgn(a_{2}^{\hcs})=\sgn(a_{2}^{\st})$ and $|a_{2}^{\hcs}| > |a_{2}^{\st}|$---the white
  noise forcing diminishes the non-Gaussian character of the vdf, thus
decreasing $\abs{a_{2}}$. Then,
just after the second jump $\chi_{1}\to\chi_{f}$ at $t=t_{w}$, despite
having the ``correct'' kinetic temperature $T_{f}$, the system is
cooling slower (faster) than at the steady state when
$a_{2}^{\hcs}-a_{2}^{\st}$, or simply $a_{2}^{\st}$, is negative
(positive), i.e. for small (large) inelasticity.

\begin{figure}
\centering
\includegraphics[width=3in]{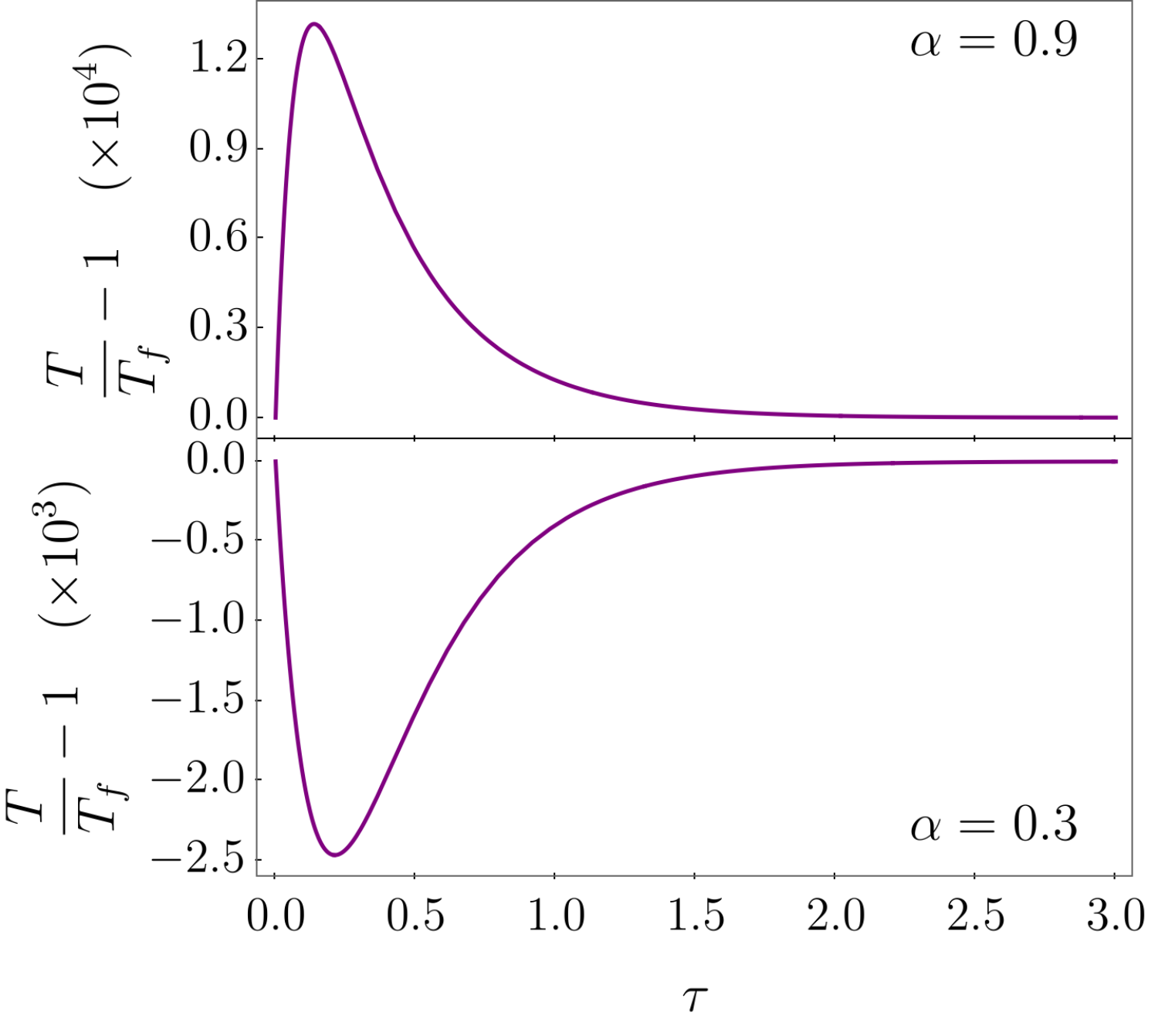}
\caption{Kovacs memory effect for the uniformly heated granular gas. The theoretical curves in the first Sonine approximation for both the normal positive hump (top panel) and the anomalous negative hump (bottom) are shown as a function of a dimensionless time $\tau$. See refs.~\cite{prados_kovacs-like_2014,trizac_memory_2014} for more details and the comparison with numerical simulations.}
\label{fig:kovacs-granular}
\end{figure}

For $t>t_{w}$, the discussion above entails that the kinetic
temperature $T(t)$ initially increases (decreases) and passes through
a maximum (minimum) before going back to $T_{f}$ when $a_{2}^{\st}<0$
($a_{2}^{\st}>0$), i.e. for small (large) inelasticity. Therefore, the
Kovacs hump is \textit{normal}, similar to that of molecular fluids---positive
and with only one maximum---for small inelasticity, whereas the Kovacs hump turns out to be \textit{anomalous},
using the term introduced in
ref.~\cite{prados_kovacs-like_2014}, for large
inelasticity: negative with one minimum.
Figure~\ref{fig:kovacs-granular} shows two examples of the
aforementioned behaviours.
In the granular gas, both the normal and the anomalous Kovacs effect
persist in the linear response regime~\cite{sanchez-rey_linear_2021}.

The Kovacs effect has also been investigated in a granular fluid of
rough particles. In addition to inelastic, collisions have a
certain degree of roughness, i.e. the tangential component of the
relative velocity is not conserved in collisions. This induces a coupling between the translational and rotational degrees of freedom.
More complex Kovacs
responses emerge, which may involve several
extrema~\cite{lasanta_emergence_2019}.

The linear response theory for molecular systems~\cite{prados_kovacs_2010} has been
generalised to athermal
systems~\cite{kursten_giant_2017,plata_kovacs-like_2017}. Specifically, the relation between the Kovacs hump and the direct
relaxation function remains valid, but the latter is not necessarily a
sum of positive modes. It is this fact that makes it possible the
emergence of the anomalous Kovacs effect, at least in linear
response~\cite{sanchez-rey_linear_2021}.

Finally, it is interesting to note that the Kovacs effect has also
been recently investigated in a variety of systems, such as active
matter~\cite{kursten_giant_2017}, disordered mechanical
systems~\cite{lahini_nonmonotonic_2017}, frictional
interfaces~\cite{dillavou_nonmonotonic_2018}, a levitated colloidal
nanoparticle~\cite{militaru_kovacs_2021}, or fluids with non-linear
drag~\cite{patron_strong_2021}.

\section{Mpemba effect}\label{sec:Mpemba}

To start with, we discuss the entropic (stochastic) Mpemba effect, triggered by the seminal Lu and Raz's
work~\cite{lu_nonequilibrium_2017}. A mesoscopic system is considered
and its time evolution is analysed in terms of the pdf  of the relevant variables, which obeys a
Markovian evolution equation (master equation, Fokker-Planck equation,
etc.) with detailed balance. Distance to equilibrium is defined in terms of a functional of
the pdf, e.g. the Kullback-Leibler divergence or other norms like the
$\mathcal{L}^{1}$ or $\mathcal{L}^{2}$ norms. By expanding the
solution of the evolution equation in the eigenfunctions of the
relevant operator, the entropic Mpemba effect is found when, under appropriate
conditions, the amplitude of the slowest relaxation mode presents a
non-monotonic dependence with the
temperature~\cite{lu_nonequilibrium_2017,klich_mpemba_2019,kumar_exponentially_2020,busiello_inducing_2021,chetrite_metastable_2021,holtzman_landau_2022,kumar_anomalous_2022,biswas_mpemba_2023}.
Also, a \textit{strong} Mpemba effect has been reported, which
arises when, by adequately choosing the system parameters, the
coefficient of the slowest relaxation mode vanishes and the relaxation
to equilibrium becomes exponentially faster
\cite{lu_nonequilibrium_2017,klich_mpemba_2019,kumar_exponentially_2020,kumar_anomalous_2022}.

In the thermal (kinetic) approach, started with Lasanta~\etal's analysis of a granular gas~\cite{lasanta_when_2017}, its
time evolution is analysed in terms of the one-particle vdf,
which evolves following a kinetic equation---Boltzmann-Fokker-Planck,
typically. The relaxation to the steady state is monitored by the
kinetic temperature.  The kinetic
approach has been employed for both granular
fluids~\cite{lasanta_when_2017,torrente_large_2019,biswas_mpemba_2020,gomez_gonzalez_mpemba-like_2021,biswas_mpemba_2021,megias_mpemba-like_2022},
in which collisions between particles are inelastic, and molecular fluids with elastic collisions but with a non-linear drag
force~\cite{santos_mpemba_2020,patron_strong_2021,megias_thermal_2022,megias_mpemba-like_2022}. The former relaxes to a NESS that is characterised by the intensity of the driving applied to balance, in average, the energy dissipated in collisions; the latter relaxes to a true equilibrium state with a Maxwellian vdf. 


{The Mpemba effect has also been studied in spin glasses within the thermal approach; it is the internal energy that displays the crossing therein~\cite{baity-jesi_mpemba_2019}. The Mpemba effect is present only in the spin-glass phase and stems from the aging dynamics of the internal energy, which is controlled by the non-equilibrium coherence length. Interestingly, this suggests that the Mpemba effect can be considered as an effective probe for the existence of a glass transition.}

There are some key differences between the stochastic and kinetic approaches. On the one hand, the monitored quantity in the kinetic approach, the kinetic
temperature {(or the energy)}, is much closer to an
experimentally measurable quantity than the abstract distance between
distributions employed in the stochastic  approach.\footnote{{For granular gases, the velocity fluctuations of an immersed rotating blade  may be used as a proxy of the kinetic temperature~\cite{scalliet_cages_2015}.}} In
addition, the thermal Mpemba effect typically takes place for short
times, far away from the final state---which makes it easier to be observed, in principle. On the other hand,  the initial conditions in the kinetic approach must
be non-stationary and thus, essentially, non-trivial to implement---although for non-linear fluids it has been discussed the aging procedure to obtain
these initial conditions, which correspond to a long-lived,
metastable, non-equilibrium state~\cite{patron_strong_2021}; whereas
the initial conditions for the entropic Mpemba effect in the stochastic  approach are  equilibrium states.\footnote{Initial stationary conditions have also
  been considered in the granular case, but an unrealistic asymmetric
  driving mechanism has to be introduced to trigger the Mpemba
  effect\cite{biswas_mpemba_2021,biswas_mpemba_2022}.}

Now we focus on the kinetic approach. Following Prados and Trizac's study of
the Kovacs effect~\cite{prados_kovacs-like_2014,trizac_memory_2014}, 
the first Sonine approximation was employed to analyse the emergence of the Mpemba effect in a granular fluid by incorporating non-Gaussianities to the picture~\cite{lasanta_when_2017}.
Indeed, it is the non-Gaussian vdf that makes the Mpemba effect possible: if it were Gaussian, the kinetic temperature  would obey a closed first-order differential equation, without additional variables, and neither the Mpemba effect nor any other memory effect would emerge.


\begin{figure}
\centering
\includegraphics[width=3in]{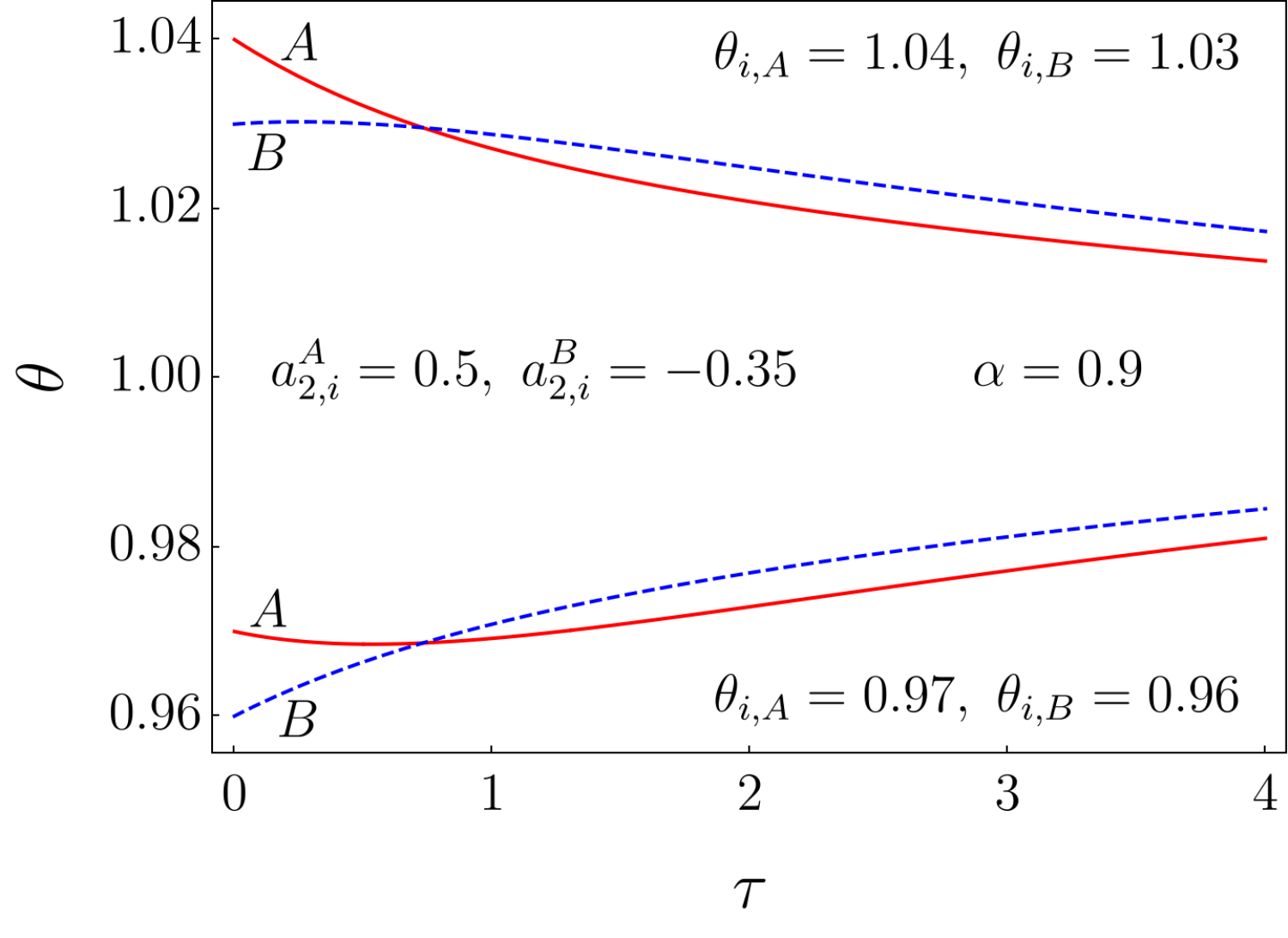}
\caption{Mpemba memory effect for the uniformly heated granular
  gas. The dimensionless temperature $\theta\equiv T/T_{\st}$ is plotted as a function of a dimensionless time $\tau$. Both the Mpemba and the inverse Mpemba effects are shown---as in fig.~\ref{fig:kovacs-granular}, only the
  theoretical curves in the first Sonine approximation. See
  ref.~\cite{lasanta_when_2017} for the comparison with numerical
  simulations.}
\label{fig:mpemba-granular}
\end{figure}
In fig.~\ref{fig:mpemba-granular}, specific examples of both the Mpemba effect, for $T_{i,A}>T_{i,B}>T_s$, and the inverse Mpemba effect, for $T_s>T_{i,A}>T_{i,B}$, are shown. The hot sample A is prepared in an initial state with kinetic
temperature $T_{i,A}$ and excess kurtosis $a_{2,i}^{A}$, and cools
down to a NESS corresponding to a certain value of the driving
$\chi_{\st}$ following the dynamical curve $T_{A}(t)$ (red solid line). The cold sample
is prepared in an initial state with kinetic temperature
$T_{i,B}<T_{i,A}$ and excess kurtosis $a_{2,i}^{B}$, and also cools
down to the same NESS following the dynamical curve $T_{B}(t)$ (blue dashed). Again, the key point is the
cooling rate  increasing with $a_{2}$: if
$a_{2,i}^{A}>a_{2,i}^{B}$, the difference of the initial cooling rates
may become large enough to facilitate the crossing of the
corresponding time evolutions $T_{A}(t)$ and $T_{B}(t)$---at least for small
enough kinetic temperature difference $\Delta T_{i}\equiv T_{i,A}-T_{i,B}$. As the initial states are not stationary states, the initial values of the kurtosis $a_{2,i}$ can be tuned to bring the Mpemba effect about.

As
the difference of the initial kurtosis
$\Delta a_{2,i}\equiv a_{2,i}^{A}>a_{2,i}^{B}$ increases, the range of
initial temperatures $\Delta T_{i}\equiv T_{i,A}-T_{i,B}$ for which
the Mpemba effect is observed increases. Since the cooling rate depends on the inelasticity $\alpha$, the range of temperatures for which the Mpemba effect emerges depends on the inelasticity as well; decreasing with it and  vanishing in the elastic
limit $\alpha\to 1$.

The thermal Mpemba effect has also been investigated for a gas of inelastic
rough hard spheres~\cite{torrente_large_2019}. Therein, the
Mpemba effect is giant, much larger than in the smooth granular gas.
The initially hotter sample may cool sooner,
even
when the initial temperatures differ by more than one order of
magnitude. The largeness of the memory effect stems from the
coupling between the translational and rotational temperatures, which
are of the same order---in the smooth case, the Mpemba
effect stemmed from the coupling with the (quite small) non-Gaussianities.

It is interesting to note that the Mpemba effect has also recently
been found in a  molecular fluid, in which the collisions between
particles are elastic, with non-linear
drag $\zeta(v)=\zeta_0 (1+ \gamma\,  m v^2/2 k_B T_s)$~\cite{santos_mpemba_2020,patron_strong_2021,megias_thermal_2022}. The non-linearity is measured by a dimensionless parameter $\gamma$, and the relevance of collisions by a dimensionless collision rate $\xi^{-1}$ ($\xi=\infty$ thus corresponds to the collisionless case.) The
kinetic temperature is not constant due to the interactions with the
thermal bath---modelled as a background fluid of particles with
comparable
mass~\cite{ferrari_particles_2007,ferrari_particles_2014,hohmann_individual_2017}. The
non-linearity of the drag implies that the evolution equation for
the temperature is coupled to higher-order cumulants of the vdf, bringing about the possible emergence of memory
effects.

\begin{figure}
\centering
\includegraphics[width=3in]{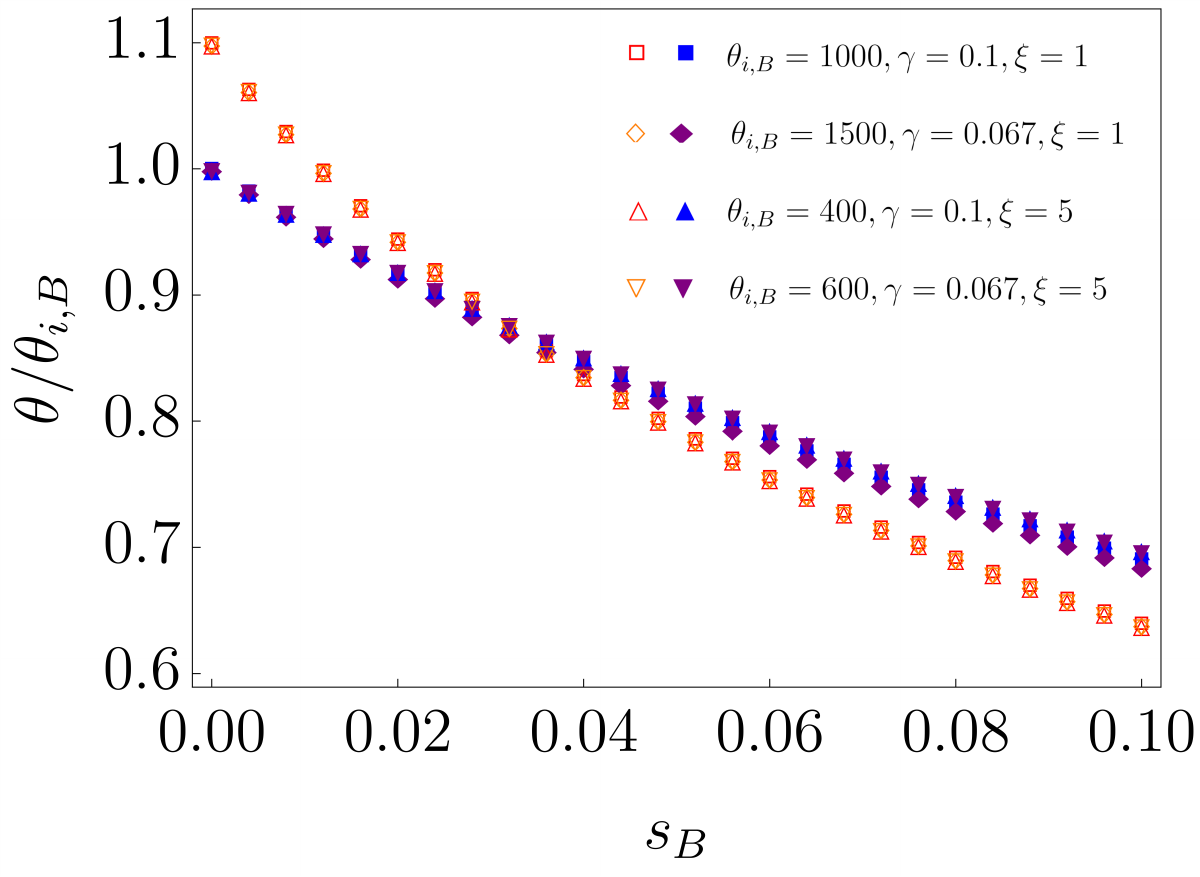}
\caption{Universal Mpemba memory effect for the molecular fluid with non-linear drag. The dimensionless temperature is again $\theta=T/T_s$. Several cold samples B {(filled blue symbols)} for different values of $(\theta_{i,B},\gamma,\xi)$ are shown together with the corresponding hot samples {A (empty red)}, for a fixed ratio of initial temperatures $\theta_{i,A}/\theta_{i,B}=1.1$. When plotted as a function of a dimensionless time $s_B=\gamma\theta_{i,B}\zeta_0 t$, all curves collapse onto a universal behaviour.}
\label{fig:univ-mpemba-nonlin}
\end{figure}

One key question, unanswered in previous studies of granular
fluids~\cite{lasanta_when_2017, torrente_large_2019}, is the aging
procedure that gives rise to the specific initial non-equilibrium
conditions that maximise the Mpemba effect. Remarkably, it is possible to give an answer for the
non-linear fluid: the hot sample must be prepared by heating it from a
much lower temperature, whereas the cold sample must be prepared by
cooling it from a much higher temperature. The high-temperature quench of the cold sample makes it
fall in a long-lived far-from-equilibrium state~\cite{patron_nonequilibrium_2023}, over which the kinetic
temperature follows a very slow 
decay to
equilibrium, which (i) increases the magnitude of the Mpemba effect and (ii) makes it universal---in the sense that the curves corresponding to different initial temperatures, non-linearity $\gamma$, and collision rate $\xi$ collapse onto a unique master curve upon a suitable rescaling, see fig.~\ref{fig:univ-mpemba-nonlin}~\cite{patron_strong_2021}.

\section{Optimal control}\label{sec:opt-control}

What is the fastest relaxation route between two given states, either
equilibrium, NESSs, or arbitrary ones? In general,
this is the problem of the brachistochrone, which has recently been
addressed in different physical
contexts~\cite{deffner_quantum_2017,plata_finite-time_2020,lam_demonstration_2021,prados_optimizing_2021,ruiz-pino_optimal_2022,patron_thermal_2022,guery-odelin_driving_2023,aghion_thermodynamic_2023,pires_optimal_2023}. It
is tempting to relate this problem with the Mpemba effect, since the
relaxation from the initially further from equilibrium state
overtaking that of the initially closer may be interpreted as the
former finding a shortcut to the common final state.

The thermal brachistochrone has been recently investigated in
uniformly driven granular
fluids~\cite{prados_optimizing_2021,ruiz-pino_optimal_2022}. It refers to the minimum time connection by controlling the intensity of
the stochastic forcing $\chi$. The protocols minimising the
connection time between the initial and final NESSs corresponding to initial and final kinetic temperatures
$T_{i}$ and $T_{f}$ are of bang-bang
type, i.e. they comprise
different time intervals in which the thermostat alternates between its
maximum and minimum available values~\cite{liberzon_calculus_2012}. 

\begin{figure}
\centering
\includegraphics[width=3in]{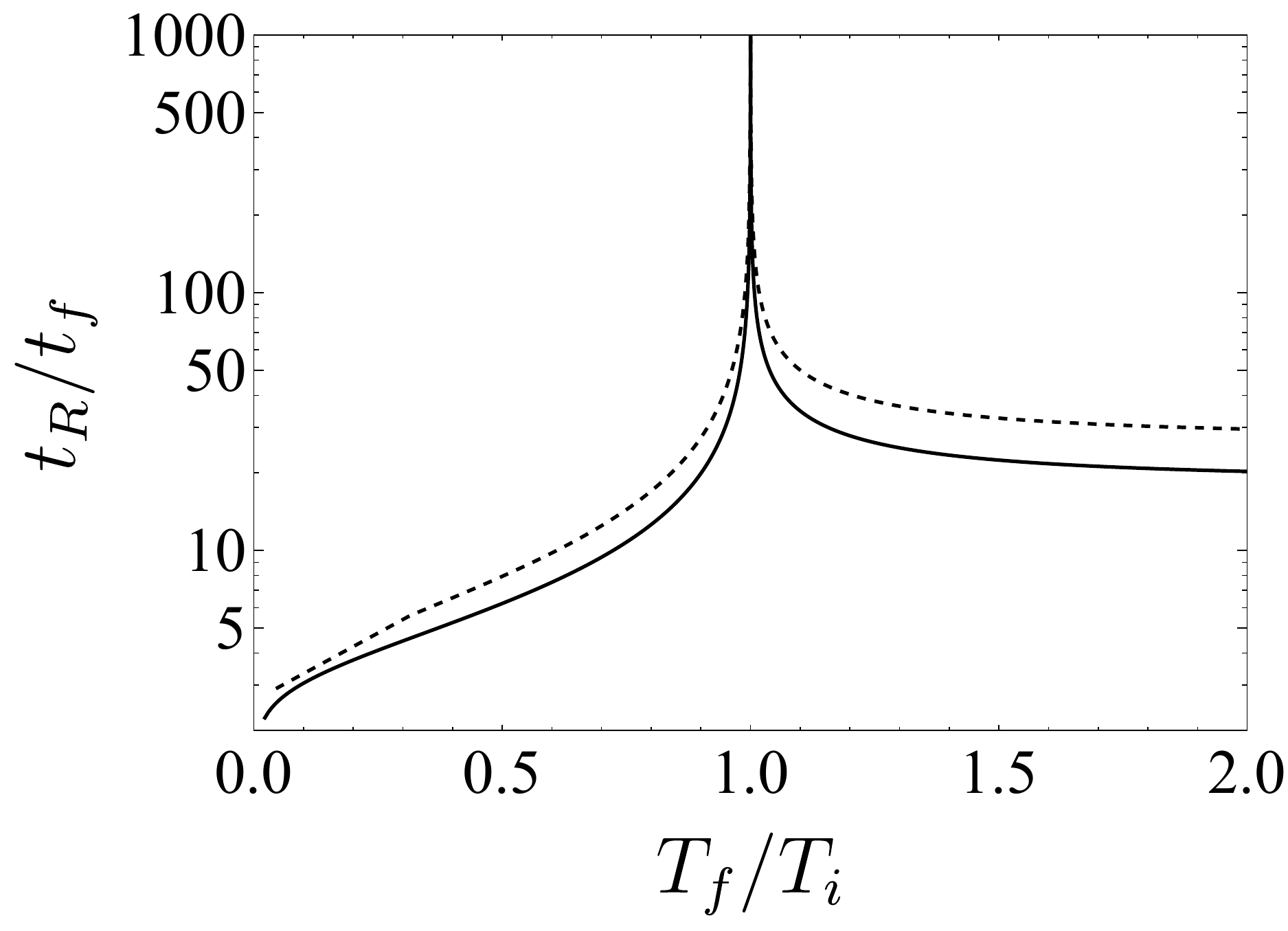}
\caption{Acceleration factor over the brachistochrone  for the granular
gas. We plot the ratio $t_R/t_{f}$ of the experimental
relaxation time to the brachistochrone time as a function of the ratio $T_{f}/T_{i}$ of the final to the initial temperature, for $\alpha=0.3$ (solid line) and $\alpha=0.8$ (dashed)---see ref.~\cite{prados_optimizing_2021} for details.
}
\label{fig:accel-factor}
\end{figure}

In the granular fluid, the time over the brachistochrone $t_{f}$
typically beats the experimental relaxation time $t_R$ by at least one order
of magnitude---see fig.~\ref{fig:accel-factor}.  Remarkably, in the usual relaxation experiment with a sudden step at $t=0$, the relaxation is never complete
in a finite time---the empirical relaxation time is defined by
estimating that the system is close ``enough'' to the final
state. On the contrary, over the brachistochrone, the system reaches exactly the final state in a finite time. 

A  similar situation, with the thermal brachistochrone given by bang-bang protocols is found in Fokker-Planck systems~\cite{prados_optimizing_2021,patron_thermal_2022}. The case of coupled harmonic oscillators that are driven from an
initial equilibrium state at temperature $T_{i}$ to a final
equilibrium state at temperature
$T_{f}$ has been analysed in detail, and an unexpected discontinuity of the minimum connection time with increasing dimension has been unveiled~\cite{patron_thermal_2022}.

\section{Discussion}\label{sec:discussion}

We have reviewed the emergence of non-equilibrium memory effects,
mainly in granular fluids and fluids with non-linear drag. Despite
being quite different from a fundamental point of view---collisions
 in granular fluids are inelastic, so they are
intrinsically out-of-equilibrium systems with non-Gaussian vdfs even in the stationary state, both types of
systems display the Kovacs and the Mpemba memory effects. Still, one key
difference between granular and non-linear fluids is the emergence of
the anomalous Kovacs effect in the former. Even when a non-linear drag
is present, the Kovacs effect is always normal when the stationary state
corresponds to equilibrium and the dynamics verify detailed balance.

The existence of these memory effects in the relaxation of the kinetic
temperature, proportional to the average kinetic energy, of granular and non-linear
fluids stems from its evolution being coupled to additional
variables, higher-order cumulants of the velocity that measure the
deviation of the vdf from the Gaussian shape. In
other words, the kinetic temperature does not suffice to univocally
determine the macroscopic state of the system. Hence, it is essential in general to keep track of the non-Gaussianities to understand the
non-equilibrium behaviour.

The thermal and entropic approaches to the Mpemba effect have been scarcely compared~\cite{megias_kinetic_2022,biswas_mpemba_2023}. In ref.~\cite{megias_kinetic_2022}, it was shown that the thermal Mpemba effect may appear without its entropic counterpart---or vice versa---in a molecular fluid with non-linear drag. Therein, some situations appear in which the kinetic temperature overshoots the stationary value, which makes it necessary to revise the usual definition of the thermal Mpemba effect in this scenario. The authors of ref.~\cite{megias_kinetic_2022}  propose a separation of the Kullback-Leibler divergence into a ``kinetic'' contribution plus a ``local-equilibrium" distribution that allows for defining a non-equilibrium temperature,  not necessarily associated with the average kinetic temperature, for any system relaxing to equilibrium. It seems worth exploring if this line of thought could lead to a unique framework for the thermal and entropic Mpemba effects.

Much progress has been made in the understanding of these memory
effects. However, there is still room for further work in this
appealing line of research. One perspective is related to their optimal
control, 
e.g. for maximising the ``positive''
consequences of a (tailored) preparation of the initial state---as in
the Mpemba effect. Therein, it seems also worth
investigating possible connections between the Mpemba effect and the
optimisation of the relaxation route to equilibrium, which has
attracted a lot of attention recently from different perspectives:
e.g. the impact of a precooling strategy~\cite{gal_precooling_2020} or the
possible asymmetry between heating and
cooling~\cite{lapolla_faster_2020,van_vu_toward_2021,ibanez_heating_2023}.








\acknowledgments
We acknowledge financial support from Grant PID2021-122588NB-I00
funded by MCIN/AEI/10.13039/ 501100011033/ and by ``ERDF A way of
making Europe'', and also from Grant
ProyExcel\_00796 funded by Junta de Andalucía's PAIDI 2020 programme. A. Patr\'on acknowledges support from the FPU programme through Grant FPU2019-4110. C. A. Plata acknowledges the funding received from EU Horizon Europe--Marie Sk\l{}odowska-Curie 2021 programme through the Postdoctoral Fellowship with ref.~101065902 (ORION). We are indebted with all the people with whom we have collaborated in
this exciting field of memory effects.

\bibliography{EPL-Perspective-memory-effects,Mi-biblioteca-01-sep-2023}






\end{document}